\documentclass[twocolumn,english,american,prx,final,superscriptaddress]{revtex4}
\usepackage{ae,aecompl}
\usepackage[T1]{fontenc}
\usepackage[latin9]{inputenc}
\usepackage{textcomp}
\usepackage{amsmath}
\usepackage{amssymb}
\usepackage{graphicx}
\usepackage{esint}

\makeatletter
\@ifundefined{textcolor}{}
{%
 \definecolor{BLACK}{gray}{0}
 \definecolor{WHITE}{gray}{1}
 \definecolor{RED}{rgb}{1,0,0}
 \definecolor{GREEN}{rgb}{0,1,0}
 \definecolor{BLUE}{rgb}{0,0,1}
 \definecolor{CYAN}{cmyk}{1,0,0,0}
 \definecolor{MAGENTA}{cmyk}{0,1,0,0}
 \definecolor{YELLOW}{cmyk}{0,0,1,0}
}

\@ifundefined{definecolor}
 {\usepackage{color}}{}
\@ifundefined{definecolor}
 {\@ifundefined{definecolor}
 {\usepackage{color}}{}
}{}

\@ifundefined{definecolor}
 {\@ifundefined{definecolor}
 {\@ifundefined{definecolor}
 {\usepackage{color}}{}
}{}
}{}
\usepackage{babel}\usepackage{longtable}\usepackage[unicode=true, pdfusetitle,
 bookmarks=true,bookmarksnumbered=false,bookmarksopen=false,
 breaklinks=false,pdfborder={0 0 1},backref=false,colorlinks=true]{hyperref}

\makeatletter
\@ifundefined{textcolor}{}
{%
 \definecolor{BLACK}{gray}{0}
 \definecolor{WHITE}{gray}{1}
 \definecolor{RED}{rgb}{1,0,0}
 \definecolor{GREEN}{rgb}{0,1,0}
 \definecolor{BLUE}{rgb}{0,0,1}
 \definecolor{CYAN}{cmyk}{1,0,0,0}
 \definecolor{MAGENTA}{cmyk}{0,1,0,0}
 \definecolor{YELLOW}{cmyk}{0,0,1,0}
 }

\usepackage{epsfig}

\usepackage{amsfonts}\usepackage{float}

\makeatother

\usepackage{babel}

\makeatother

\usepackage{babel}

\makeatother

\usepackage{babel}

\makeatother

\makeatother

\usepackage{babel}

\makeatother

\usepackage{babel}

\makeatother

\usepackage{babel}
\begin{document}

\title{Maximizing information on the environment by dynamically controlled
qubit probes}

\author{Analia Zwick}

\affiliation{Weizmann Institute of Science, Rehovot 76100, Israel}

\author{Gonzalo A. \'Alvarez}

\affiliation{Weizmann Institute of Science, Rehovot 76100, Israel}

\author{Gershon Kurizki}

\affiliation{Weizmann Institute of Science, Rehovot 76100, Israel}
\begin{abstract}
We explore the ability of a qubit probe to characterize unknown parameters
of its environment. By resorting to quantum estimation theory, we
analytically find the ultimate bound on the precision of estimating
key parameters of a broad class of ubiquitous environmental noises
(``baths'') which the qubit may probe. These include the probe-bath
coupling strength, the correlation time of generic bath spectra, the
power laws governing these spectra, as well as their dephasing times
$T_{2}$. Our central result is that by optimizing the dynamical control
on the probe under realistic constraints one may attain the maximal
accuracy bound on the estimation of these parameters by the least
number of measurements possible. Applications of this protocol that
combines dynamical control and estimation theory tools to quantum
sensing are illustrated for a nitrogen-vacancy center in diamond used
as a probe.
\end{abstract}


\maketitle

\section{Introduction}

Controlled spin-$\frac{1}{2}$ particles (qubits) are sensitive probes
of the structure and properties of highly complex molecular, atomic
or solid-state quantum systems. Novel quantum technologies requiring
such high sensitivity at the nanoscale are based on qubit probes serving
as sensors \cite{Taylor2008,Balasubramanian2008,Kucsko2013,Neumann2013,Toyli2013,Steinert2013,Grinolds2014,Sushkov2014}
or monitors of biological or chemical processes \cite{Mittermaier2006,Smith2012,Alvarez2013}.
Here we focus on \emph{the extraction of information characterizing
the environment of a qubit probe}, by monitoring the decoherence process
that the qubit undergoes \cite{Zurek2003}. Dynamical control, originally
conceived as a tool for reducing decoherence effects \cite{Viola1999,Viola1999b,kofman_universal_2001,kofman_unified_2004,gordon_universal_2007,Khodjasteh2005,Souza2012},
has also been shown to be a valuable source of information on environmental
noises \cite{Meriles2010,Almog2011,Bylander2011,Alvarez2011,Smith2012,alvarez_controlling_2012,Cywinski2014}.
This information is revealed by the dependence of the decoherence
rate of the qubit probe on a control-field parameter, owing to the
fact that, under weak coupling of the qubit and the environment (``bath''),
this rate is universally expressed by the overlap of the environmental
noise spectrum and a spectral filter function that is solely determined
by the dynamical control \cite{kofman_universal_2001,kofman_unified_2004,gordon_universal_2007,clausen_bath-optimized_2010}.
The filter function can then be designed by varying the control field
to scan the noise spectrum. The information obtained from this procedure,
dubbed ``noise spectroscopy'' \cite{Alvarez2011,Bylander2011},
is not only essential for designing the most effective (optimal) dynamical
protection from decoherence caused by a given environment \cite{gordon_optimal_2008,Uys2009,clausen_bath-optimized_2010,KLV10,Ajoy2011,Hayes2011,Calarco_speeding_2013,Khodjasteh2013}
of quantum-information processing \cite{gordon_Dynamical_2008,vio09,paz-silva_zeno_2012,Kabytayev2014},
quantum-state transfer \cite{alvarez_perfect_2010,Zwick_Optimized_2014},
Hamiltonian engineering \cite{alvarez_perfect_2010,alvarez_controlling_2012,ajoy_quantum_2013}
and quantum-state storage \cite{petrosyan_reversible_2009,bensky_optimizing_2012}.
It may also become a means of understanding physical or chemical processes
\cite{Mittermaier2006,Smith2012,alvarez_controlling_2012,Alvarez2013}
by analyzing their noise fluctuations in magnetic resonance spectroscopy
and imaging with nanoscale resolution \cite{Alvarez2013,Alvarez2013_JMR,Mamin2013,Staudacher2013,Steinert2013,Grinolds2014,Sushkov2014}.

In order to further advance this promising noise spectroscopy and
broaden its applicability, it is imperative to find \emph{the best
general strategy} for extracting the environmental (``bath-induced'')
noise-spectrum information from the qubit-probe decoherence. The strategy
we adopt aims at minimizing the error in estimating \emph{key parameters}
of the noise (bath) spectrum by measurements performed on the qubit
probe (Sec.\,\ref{sec:Dependence-of-the}). The minimal error is
determined by the maximal quantum Fisher information (QFI) \cite{Caves_1994_fisher,Paris2009_QUANTUM-ESTIMATION,escher2011general,Paris2014_Characterization-of-classical-Gaussian}
gathered by measurements in the optimal basis (Sec.\,\ref{sec:Optimal-estimation-under}).
We here find (Sec.\,\ref{sub:Ultimate-estimation-bounds}) the ultimate
error bounds for unbiased estimators\emph{ }when the qubit probe,
under arbitrary control, undergoes pure dephasing in the probe-bath
weak-coupling regime. Motivated by practical experimental considerations
and constraints, these ultimate bounds attain the best estimation
precision by the least number of measurements possible. We achieve
these goals by \emph{replacing the free-evolution} (-induction) decay
(FID) of the qubit state prior to each measurement by\emph{ dynamically-controlled
evolution} designed to ensure the convergence to the ultimate error
bound for the particular bath-spectrum parameter to be estimated (Sec.\,\ref{sub:Key-parameter-estimation}).
For each such parameter, an appropriate filter-function must be generated
by dynamical control (Sec.\,\ref{sub:Achieving-the-ultimate-bound})
\cite{gordon_universal_2007,clausen_bath-optimized_2010}. We here
focus on determining the general conditions that have to be satisfied
for designing the filter function to attain the ultimate bounds. 

The first step in the proposed strategy is the estimation of the coupling
strength $g$ of the bath to the probe, that can be interpreted as
the noise variance. The proposed appropriate filter function is generated
by \emph{projections} onto an eigenstate of the $\sigma_{x}$ probe-operator
at a rate that conforms to the quantum Zeno regime \cite{kofman_acceleration_2000,kofman_universal_2001,alvarez_controlling_2012}
(Sec.\,\ref{sub:Estimating-the-probe-bath}). This procedure, which
\textit{does not require prior knowledge of the bath spectral lineshape},
has been experimentally exploited to determine the coupling strengths
of complex spin-networks without maximizing or optimizing the information
obtained \cite{alvarez_controlling_2012}.\emph{ }

Once the coupling strength $g$ is optimally estimated, the bath spectra
have to be characterized by their normalized lineshapes. These spectra
crucially depend on the bath correlation time $\tau_{c}$ often unknown
to us: it is typically the inverse of $\omega_{c}$, which is the
width or the cutoff of the bath spectrum. We show that convergence
to the lowest error bound on $\tau_{c}$ is achievable through filter
functions generated by common types of coherent dynamical-control
sequences (Sec.\,\ref{sub:Estimating-the-correlation}). By contrast,
free-induction decay (FID) of the qubit coherence $\left\langle \sigma_{x}(t)\right\rangle $
may preclude such convergence. 

We further show for a family of generic baths that \textit{optimized
convergence requires a filter function that only samples (overlaps)}
a \emph{power-law region} of the bath spectrum (Sec.\,\ref{sub:Key-parameter-estimation}).
Such spectral features \emph{characterize omnipresent baths}: sub-Ohmic,
Ohmic and super-Ohmic baths whose spectra obey a power-law at low
frequencies, as well as noise spectra of generalized Ornstein-Uhlenbeck
processes characterized by a power-law tail at high frequencies. These
types of bath spectra are ubiquitous in solid-state, liquid or gas
phases \cite{Alvarez2011,Bar-Gill2012,Medford2012,Alvarez2013,Romach2014}
where they represent collisional or diffusion processes \cite{Suter1985,Feintuch2004}.
Other environmental parameters, such as spectral power-law exponents,
the $T_{2}$ decoherence time and diffusion coefficients, are shown
to obey bounds analogous to those of $g$ and $\tau_{c}$ (Sec.\,\ref{sub:Power-law-estimation}). 

Finally, we demonstrate the practical feasibility of experiments that
may attain the ultimate precision bounds by resorting to a real-time
adaptive estimation protocol based on a Bayesian estimator and an
online experimental learning design \cite{Cory2012_Robust-online-Hamiltonian-learning,Sergeevich_Characterization_2011,Yacoby2014,Cappellaro_2012_Spin-bath}
(Sec.\,\ref{sec:Real-time-estimation-protocol}). We resort to this
protocol to illustrate the ability to achieve the predicted analytical
bounds in an efficient way under experimentally relevant conditions,
e.g. for nitrogen-vacancy center (NVC) probes in diamond \cite{Taylor2008,Balasubramanian2008,Kucsko2013,Neumann2013,Toyli2013,Steinert2013,Grinolds2014,Sushkov2014,Staudacher2013,Bar-Gill2012}.
Thus, the present analytical theory, supported by adaptive-estimation
simulations, suggests that the proposed \emph{synthesis of optimally-controlled
noise spectroscopy and estimation theory} can become a powerful, broadly
applicable, diagnostic tool (Sec.\,\ref{sec:Discussion}).

\section{\noindent Dependence of the qubit-probe dephasing on an unknown environmental
parameter\label{sec:Dependence-of-the}}

\begin{figure*}
\includegraphics[width=0.85\textwidth]{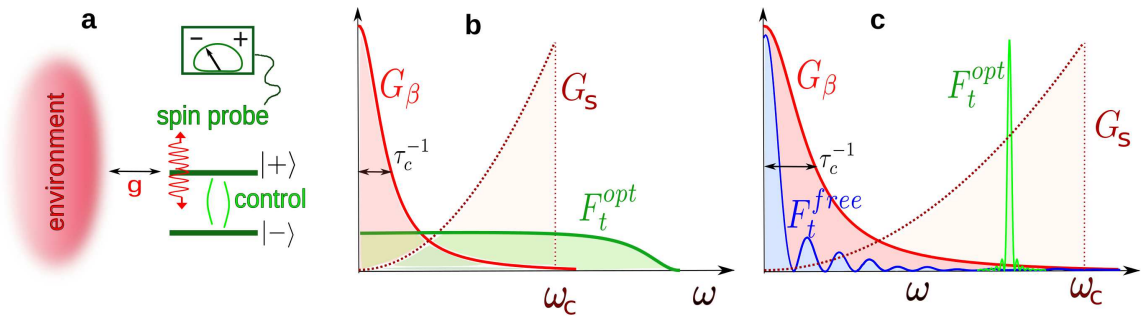}

\protect\caption{\label{fig1_scheme} Scheme of noise-spectra parameter estimation.
(a) Scheme of the probing process and its dynamical control: Dynamically
controlled qubit-probe undergoes dephasing by the environment (bath).
(b) An optimal qubit-probe filter function $F_{t}^{opt}(\omega)$
generated by frequent projections is spectrally flat to the extent
that it conforms to be quantum Zeno regime and allows the optimal
determination of the coupling strength $g$ regardless of the shape
of $S(\omega)$.\textbf{ }(c) A filter function of an optimally controlled
qubit-probe, $F_{t}^{opt}(\omega)$, for obtaining the ultimate bound
on the correlation time ($\tau_{c}$) estimation. Such a filter must
overlap only with the power-law tail of the Ornstein-Uhlenbeck noise
spectrum $S_{\beta}(\tau_{c},\omega)$, or with $\omega>0$ but not
with $\omega_{c}$ for the (super) Ohmic spectrum $S_{s}(\tau_{c},\omega)$.
The free-evolution (FID) filter function $F_{t}^{free}(\omega)$ does
not fulfill the requirements for achieving the bound, since it is
centered at $\omega=0$.}
\end{figure*}

\noindent We consider a qubit-probe that experiences proper-dephasing
by the environment (bath) under the action of the system-bath interaction
Hamiltonian

\noindent 
\begin{equation}
H_{SB}=g\sigma_{z}B,\label{eq:1}
\end{equation}
where $g$ is the probe-bath interaction strength, $\sigma_{z}$ is
the appropriate Pauli operator for the probe and $B$ is the bath
operator (Fig. \ref{fig1_scheme}a). To obtain maximal information
on the environment, a convenient initial probe-state is the symmetric
superposition of the qubit-up/-down states in the $\sigma_{z}$ basis,
\begin{equation}
\frac{1}{\sqrt{2}}(\left|\uparrow\right\rangle +\left|\downarrow\right\rangle )=\left|+\right\rangle ,
\end{equation}
and the optimal observable for the probe-state measurement is $\sigma_{x}$
\cite{Paris2014_Characterization-of-classical-Gaussian} (Appendix
\ref{sub:Quantum-Fisher-information}). Our goal is to determine the
dynamical control efficacy for estimating, one by one, the unknown
parameters such as $g$ and $\tau_{c}$ that characterize the coupling
bath spectrum.

We shall denote the particular bath parameter by $x_{B}$, and, accordingly,
the bath coupling spectrum by $G(x_{B},\omega)$. Under proper dephasing,
\begin{equation}
\left\langle \sigma_{x}(x_{B},t)\right\rangle =Tr(\rho_{S}(x_{B},t)\sigma_{x})=e^{-\mathcal{J}(x_{B},t)},
\end{equation}
where $\mathcal{J}(x_{B},t)$ is the attenuation factor due to dephasing.
In \emph{the} \emph{probe-bath weak-coupling regime} known as the
Born approximation for the bath, wherein the qubit-probe negligibly
influences the environment \cite{Breuer2007}, it obeys the universal
formula (Appendix \ref{sub:Quantum-probe-under}) \cite{kofman_universal_2001,kofman_unified_2004,gordon_universal_2007,clausen_bath-optimized_2010}
\begin{equation}
\mathcal{J}(x_{B},t)=\int_{-\infty}^{\infty}d\omega F_{t}(\omega)G(x_{B},\omega),\label{eq:Dephasing rate xB}
\end{equation}
where $F_{t}(\omega)$ is a filter function which explicitly depends
upon the dynamical control of the qubit-probe during time $t$.

This universal formula is exact for Gaussian noise. Yet it applies
to any noise under the weak-coupling assumption in Eq. (\ref{eq:Dephasing rate xB}).
In our bath-optimized control theory \cite{kofman_universal_2001,kofman_unified_2004,gordon_universal_2007,gordon_optimal_2008,clausen_bath-optimized_2010},
the filter function is obtained to ensure optimal control for any
given bath and task at hand. This theory does not impose any requirement
on the temporal shape of dynamical control that defines the filter
function: the control may be continuous or pulsed, coherent or projective,
in contrast to the dynamical-decoupling formulation \cite{Viola1999b,Khodjasteh2005,KLV10}.
We stress that the optimal filter function obtained by this bath-optimized
control theory does not require high-order corrections as opposed
to dynamical decoupling control \cite{clausen_bath-optimized_2010,gordon_optimal_2008}.

The information about the unknown bath parameter $x_{B}$ is encoded
in the protocol defined by Eqs. (\ref{eq:1})-(\ref{eq:Dephasing rate xB})
(Fig. \ref{fig1_scheme}a) by the probabilities $p$ of finding the
qubit in the $\left|+\right\rangle $ (symmetric) or $\left|-\right\rangle $
(antisymmetric) state when measuring $\sigma_{x}$. These probabilities
obey
\begin{equation}
p(\pm|x_{B},t)=\frac{1}{2}\left(1\pm e^{-\mathcal{J}(x_{B},t)}\right).\label{eq:p+-}
\end{equation}

\section{\noindent \textit{\emph{Optimal estimation under dynamical control}}}

\noindent \textbf{\textit{\emph{\label{sec:Optimal-estimation-under}}}}The
minimum achievable relative error of the \emph{unbiased} estimation
of a single unknown parameter $x_{B}$ is determined by the quantum
Cramer-Rao bound to be
\begin{equation}
\varepsilon(x_{B},t)=\frac{\delta x_{B}}{x_{B}}\geq\frac{1}{x_{B}\sqrt{N_{m}\mathcal{F_{Q}}(x_{B},t)}},\label{eq:Relative-Error}
\end{equation}
where $N_{m}$ is the number of measurements and $\mathcal{F_{Q}}(x_{B},t)$
is the quantum Fisher information (QFI) that quantifies the maximum
amount of information on $x_{B}$ that can be extracted from a given
state \cite{Caves_1994_fisher}. Therefore, we set out to maximize
the information 
\begin{equation}
\mathcal{F_{Q}}(x_{B},t_{opt})=\underset{t}{max}\mathcal{F_{Q}}(x_{B},t),
\end{equation}
by choosing the optimal time $t_{opt}$ to perform the measurement
and an appropriate dynamical-control scheme (prior to the measurement),
to obtain the minimal attainable (relative) error in the estimation
of $x_{B}$ for a given bath spectrum. For a qubit-probe obeying Eq.
(\ref{eq:p+-}), the QFI is gathered by measurements in the optimal
basis (Appendix \ref{sub:Quantum-Fisher-information})
\begin{equation}
\mathcal{F_{Q}}(x_{B},t)=\frac{e^{-2\mathcal{J}(x_{B},t)}}{1-e^{-2\mathcal{J}(x_{B},t)}}\left(\frac{\partial\mathcal{J}(x_{B},t)}{\partial x_{B}}\right)^{2}.\label{eq:QFI_R}
\end{equation}

\begin{figure*}
\includegraphics[width=0.7\textwidth]{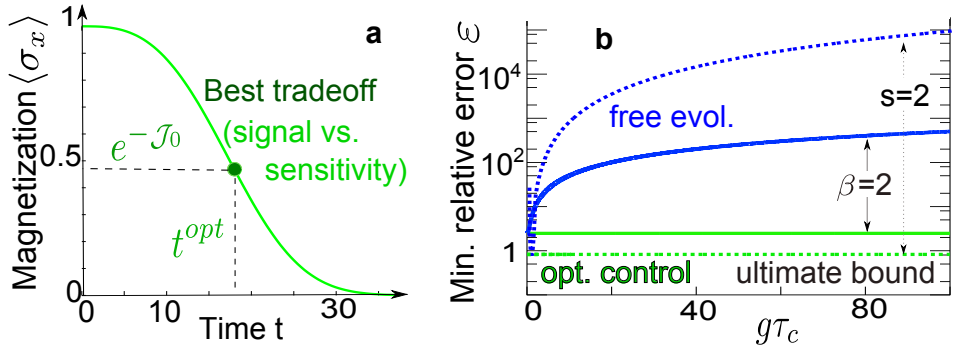}

\protect\caption{\label{fig2_opt-cntrl_Lorentz}Optimization for achieving the ultimate
precision bound. (a) Magnetization $\langle\sigma_{x}\rangle$ as
a function of the measurement time $t$ of a qubit-probe experiencing
dephasing due to Ornstein-Uhlenbeck process, with Lorentzian spectrum
$G_{\beta=2}(\tau_{c}\!=\!10,\omega)$ with $g=1$ under a CPMG control
sequence with $N\!=\!8$ pulses using the exact analytical expression
derived from Eq. (\ref{eq:Dephasing rate xB}) \cite{Alvarez2013}.
The optimal time $t^{opt}$ that determines the best tradeoff between
the signal contrast (almost halfway between 0 and 1) and the highest
sensitivity of the decay rate to $\tau_{c}$, provides the most accurate
estimation of $\tau_{c}$. (b) Minimal relative error $\varepsilon(\tau_{c},t^{opt})$
per measurement ($N_{m}\!=\!1)$ of the estimation of $\tau_{c}$
as a function of $g\tau_{c}$ for a Lorentzian spectrum $G_{\beta=2}(\tau_{c}=10,\omega)$
(dashed lines) and a super-ohmic spectrum $G_{s=2}(\tau_{c}=10,\omega)$
(solid lines) calculated from the integral of Eq. (\ref{eq:Dephasing rate xB}),
both with $g=1$. The minimal relative error under optimal control
(CPMG/CW with $N$ pulses/cycles satisfying the conditions given in
the main text, green lines) achieves the ultimate bounds \emph{($\frac{\varepsilon_{0}}{3\sqrt{N_{m}}}$}
(dashed-) and \emph{$\frac{\varepsilon_{0}}{\sqrt{N_{m}}}$} (solid-)
green lines for $s=2$ and $\beta=2$ respectively). It can improve
the best precision obtainable under free-evolution (blue lines) by
several orders of magnitude. }
\end{figure*}

From this expression, the QFI maximum is obtained by finding the optimal
tradeoff between the signal-amplitude contrast (the magnitude of $\frac{e^{-2\mathcal{J}}}{1-e^{-2\mathcal{J}}}$)
and the sensitivity of the signal-attenuation to the parameter $x_{B}$
(the derivative $\frac{\partial\mathcal{J}}{\partial x_{B}}$). Obviously,
neither should be too small if Eq. (\ref{eq:QFI_R}) is to be maximized.
Their optimal tradeoff determines $t_{opt}$ for a given control scheme
(Fig. \ref{fig2_opt-cntrl_Lorentz}a). To evaluate the efficacy of
different dynamical controls in attaining the highest accuracy in
$x_{B}$, we then use the error (\ref{eq:Relative-Error}) at $t_{opt}$,
\emph{i.e.} $\varepsilon(x_{B},t_{opt})$, as the figure of merit.

We focus here on the ultimate error bounds for the estimation precision
\textit{per measurement} for the qubit-probe that undergoes pure dephasing
within the weak-coupling regime. These bounds are of practical importance
in typical experimental situations where the initialization and readout
times, normally determined by $T_{1}\gg T_{2}$, constrain the time
interval between measurements.

Alternatively, in instances where $T_{1}$ is not the dominant constraint,
one may be interested in the ultimate precision bound attainable during
a given interrogation time that extends over many measurements, $T=N_{m}t$.
In such cases the Fisher information \textit{per unit time} has to
be maximized: 
\begin{equation}
\mathcal{F_{Q}}(x_{B},t_{opt})/t_{opt}=\underset{t}{max}(\mathcal{F_{Q}}(x_{B},t)/t).
\end{equation}

\subsection{\noindent Ultimate estimation bounds\label{sub:Ultimate-estimation-bounds}}

As discussed below, a \textit{\emph{broad class of practically relevant
environmental }}noise processes\textit{\emph{ }}cause the attenuation
factor $\mathcal{J}(x_{B},t)$ to have a power-law dependence in $x_{B}$
with an exponent $\alpha$, where the derivative of $\mathcal{J}(x_{B},t)$
is tightly bounded by its value in the power-law region, as
\begin{equation}
\left|\frac{\partial\mathcal{J}(x_{B},t)}{\partial x_{B}}\right|\le\frac{\alpha\mathcal{J}(x_{B},t)}{x_{B}}.\label{eq: deriv-Att-fact_x_B}
\end{equation}
The equality strictly holds when \foreignlanguage{english}{\emph{$\mathcal{J}(x_{B},t)$}}\textit{\emph{
is an }}\textit{homogeneous function of degree}\emph{ }$\alpha$ for
the parameter $x_{B}$.

\noindent \textit{\emph{The bound (}}\ref{eq: deriv-Att-fact_x_B}\textit{\emph{)
leads to a}} tight lower bound on the relative error in (\ref{eq:Relative-Error}),
which holds\emph{ for either free decay (FID) or }a\emph{rbitrary
dynamical control} of the probe coherence\textit{\emph{ }}(Appendix
\ref{sub:Bound-derivation})
\begin{equation}
\varepsilon(x_{B},t)\geq\frac{\sqrt{1-e^{-2\mathcal{J}_{0}}}}{\alpha\sqrt{N_{m}}\,\mathcal{J}_{0}\, e^{-\mathcal{J}_{0}}}=\frac{\varepsilon_{0}}{\alpha\sqrt{N_{m}}}.\label{eq:ultimate bound}
\end{equation}

\noindent Here the value of $\mathcal{J}=\mathcal{J}{}_{0}$ that
minimizes this bound, $\mathcal{J}_{0}\!=\!1\!+\!\frac{1}{2}W(-2e^{-2})\!\approx\!0.8$,
$W(z)$ being the Lambert function, yields the equality in (\ref{eq:ultimate bound}),
with $\varepsilon_{0}\thickapprox2.48$. We have thus obtained a \emph{universal,
ultimate error bound for the unbiased estimation of $x_{B}$ in a
broad class of noise spectra} probed by a qubit-probe undergoing dephasing
within the probe-bath weak-coupling regime.

\subsection{\noindent Key parameter estimation\textit{\label{sub:Key-parameter-estimation}}}

This general bound applies to the estimation of any parameters that
satisfy $\mathcal{J}(x_{B},t)\propto x_{B}^{\pm\alpha}$. We here
discuss in detail some examples of estimating key parameters that
belong to this class in typical scenarios of practical importance:\emph{}\\
\emph{(i)} The effective coupling strength $g$ of the qubit-probe
with the dephasing bath is required for defining \foreignlanguage{english}{
\begin{equation}
G(x_{B}\!=\! g,\omega)=g^{2}S(\omega),
\end{equation}
$S(\omega)$ being the normalized spectral density of the bath autocorrelation
function}, so that \foreignlanguage{english}{$G(g,\omega)$ and $\mathcal{J}(g,t)$}
are \textit{homogeneous functions of degree}\emph{ }$\alpha=2$ in
$g$\emph{.}\\
\emph{(iia)} The correlation time $\tau_{c}$ of noise fluctuations\textit{\emph{
is required for describing generalized}} Ornstein-Uhlenbeck\textit{\emph{
processes, with normalized spectral densities}}
\begin{equation}
S_{\beta}(x_{B}\!=\!\tau_{c},\omega)=\mathcal{A}_{\beta}\frac{\tau_{c}}{1+\omega^{\beta}\tau_{c}^{\beta}},\label{eq:power-law tail noise spectrum}
\end{equation}

\noindent where $\beta\geq2$ is an even integer and $\mathcal{A}_{\beta}=\frac{\beta}{2\pi}sin(\frac{\pi}{\beta})$
is the normalization factor.\\
\emph{(iib)} The correlation time $\tau_{c}=\omega_{c}^{-1}$, that
is the inverse of the cutoff frequency $\omega_{c}$, in generalized
Ohmic spectra
\begin{equation}
S_{s}(x_{B}\!=\!\tau_{c},\omega)=(s+1)\omega_{c}^{-(s+1)}\omega^{s}\Theta(\omega)\Theta(\omega-\omega_{c}),\label{eq: ohmic power-law noise spectrum}
\end{equation}

\noindent where $\Theta(\omega)$ is the step function, $s=1$ stands
for a Ohmic spectrum, $0<s<1$ for a sub-Ohmic spectrum, and $s>1$
for its super-Ohmic counterpart. 

\noindent Both spectral densities \textit{\emph{(}}\ref{eq:power-law tail noise spectrum}\textit{\emph{)}}
and (\ref{eq: ohmic power-law noise spectrum}) satisfy\textit{\emph{
$\left|\frac{\partial S}{\partial\tau_{c}}\right|\leq\frac{\alpha S}{\tau_{c}}$,
and consequently the bound (}}\ref{eq: deriv-Att-fact_x_B}\textit{\emph{).
Furthermore, both spectra attain the ultimate bound (}}\ref{eq:ultimate bound}\textit{\emph{)
at frequency ranges where they have power-law dependence (Appendix
}}\ref{sub:Bound-derivation}\textit{\emph{): spectral density (}}\ref{eq:power-law tail noise spectrum}\textit{\emph{)
at high frequencies, where it becomes a homogeneous function of degree}}
\emph{$\alpha=\beta-1$,}\textit{\emph{ and spectral density (}}\ref{eq: ohmic power-law noise spectrum}\textit{\emph{)
at low frequencies, when }}we restrict ourselves to \textit{\emph{$\omega<\omega_{c}$
to avoid the cutoff}} \textit{\emph{effects, thus rendering the spectral
density a homogeneous function of degree $\alpha=s+1$.}}

\noindent \emph{(iii)} The dephasing time $T_{2}$ in the attenuation
exponent which is a homogeneous function of degree $\mbox{\ensuremath{\alpha}}$,
$\mathcal{J}(x_{B}=T_{2},t)=\left(t/T_{2}\right)^{\alpha}$, may be
estimated down to the ultimate bound.

\section{\noindent \textit{\emph{Dynamical control strategies for achieving
the ultimate bound\label{sec:Dynamical-control-strategies}}}}

\subsection{\noindent Achieving the ultimate-bound by dynamical control\label{sub:Achieving-the-ultimate-bound}}

\noindent It follows from the discussion above that the ultimate error
bound, Eq. (\ref{eq:ultimate bound}), in the estimation of $x_{B}=g$
or $\tau_{c}$, can be attained provided that the dynamical control
on the probe generates a filter $F_{t}(\omega)$ that extends only
over the frequency band where the noise spectrum behaves as a power-law
in $x_{B}^{\pm\alpha}$ and the equality in Eq. (\ref{eq: deriv-Att-fact_x_B})
is fulfilled. Then, upon adjusting the total control time such that
$\mathcal{J}(x_{B},t^{opt})=\mathcal{J}_{0}$, the equalities in Eq.
(\ref{eq:ultimate bound}) are also fulfilled.\textit{\emph{ }}

\noindent \textit{\emph{We note that these are general conditions
for attaining the ultimate error bound per measurement, but they do
not invoke the optimization of the filter function under specific
constraints that may be imposed in a given experimental setup. In
such setups further optimization is required to approach as best we
can the ultimate error bounds.\vspace{-3mm}
}}

\subsection{\noindent $\!\!$Estimating the probe-bath interaction strength $\! g$\label{sub:Estimating-the-probe-bath}}

\noindent If $S(\omega)$ is \emph{unknown} apart from a crude estimate
of its overall width, dynamical control is needed for estimating $g$
down to the ultimate precision bound. The appropriate control is such
that the filter function $F_{t}(\omega)=F_{t}$ is flat in the domain
of $S(\omega)$, leading to an attenuation factor $\mathcal{J}=F_{t}g^{2}$
that is independent of the noise spectrum. For free evolution this
limit holds when the interval is much shorter than the correlation
time $\tau_{c}$ of the environment, so that the qubit-probe evolves
freely (Appendix \ref{sub:Dephasinh-Rate}). This condition can only
be fulfilled for large enough $\tau_{c}$.

To overcome this limitation, we may instead realize such a filter
at times larger than $\tau_{c}$ via repeated stroboscopic projections
of the qubit-probe in the basis of $\left|\pm\right\rangle $, \emph{i.e.}
by quantum non-demolition (QND) measurements at a rate that is high
enough to conform to the quantum Zeno regime \cite{kofman_acceleration_2000}.
The filter function then describes spectral broadening of the $\left|\pm\right\rangle $
eigenvalues far beyond the width of $S(\omega)$ (Fig. \ref{fig1_scheme}b).
The attenuation factor is then \cite{kofman_universal_2001,gordon_universal_2007,alvarez_controlling_2012}
(Appendix \ref{sub:Dephasinh-Rate})
\begin{equation}
\mathcal{J}_{Zeno}(g,t)=\frac{g^{2}t^{2}}{2N},\textrm{ if }\frac{t}{N}\ll\tau_{c},\label{eq:Att factor J_Zeno}
\end{equation}
obtained for $N$ QND measurements during the total control time $t$.
The advantage of this Zeno regime is that it requires \foreignlanguage{english}{$\frac{t}{N}\ll\tau_{c}$,
rather than} the total time $t$, to be less than $\tau_{c}$. Since
the outcomes of these measurements \emph{need not be read out}, but
rather used to guide the evolution, they may \emph{be emulated} by
impulsive noise-induced dephasing of the qubit probe that has the
same effect as a projection on the probe evolution \cite{Alvarez2010,alvarez_controlling_2012,gordon_universal_2007}.
This QZE regime has been already exploited experimentally to determine
the coupling strengths of complex spin-networks \cite{alvarez_controlling_2012}.

Under the quantum-Zeno condition of (\ref{eq:Att factor J_Zeno}),
the error estimation bound (\ref{eq:ultimate bound}) is achieved
when the total control time (after which $\sigma_{x}$ is measured
and read out - see (\ref{eq:p+-})) is chosen to have the optimal
value
\begin{equation}
t_{Zeno}^{opt}=\frac{\sqrt{2N\mathcal{J}_{0}}}{g},\mathrm{\textrm{ provided that }\mbox{\ensuremath{N\gg\frac{2\mathcal{J}_{0}}{g^{2}\tau_{c}^{2}}}}.}\label{eq:toptZeno}
\end{equation}
This equation (further discussed in Appendix \ref{sub:Dephasinh-Rate})
constitutes the main condition on an optimal filter designed to estimate
$g$.

\subsection{\noindent Estimating the correlation time $\tau_{c}$ \label{sub:Estimating-the-correlation}}

\noindent To achieve the ultimate precision bound (\ref{eq:ultimate bound})
for the estimation of $\tau_{c}$ of noise spectra (\ref{eq:power-law tail noise spectrum})-(\ref{eq: ohmic power-law noise spectrum}),
the control on the probe should generate a filter $F_{t}(\omega)$
that only overlaps with the power-law portion of $S_{\beta(s)}(\tau_{c},\omega)\propto\tau_{c}^{\mp\alpha}\omega^{\mp(\alpha\pm1)}$,
as shown in Fig. \ref{fig1_scheme}c. By contrast, FID of the probe
coherence generates a filter (sinc) function centered at zero frequency,
thus \emph{preventing the bound} in Eq. (\ref{eq:ultimate bound})
\emph{from being reached}.

\noindent While various controls may allow the best estimation according
to Eq. (\ref{eq:ultimate bound}), we here analytically study the
conditions for achieving the ultimate bound under standard CPMG sequences
of $\pi$ pulses \cite{Slichter1990}, or under \emph{continuous-wave
driving} (CW). In control sequences of equidistant pulses $N\gg1$,
the filter function $F_{t}(\omega)$ converges to a sum of delta functions
(narrowband filters) centered at the harmonics of the inverse CPMG
time period, while for CW there is a single frequency component \cite{gordon_universal_2007,Alvarez2011,Ajoy2011}
(Appendix \ref{sub:Dephasinh-Rate}). In the following we use these
filter functions, to analytically infer the required total control
time $t$ and $N$ that allow the bound to be attained for the two
classes of power-law spectra in Eqs. \textit{\emph{(}}\ref{eq:power-law tail noise spectrum}\textit{\emph{)}}
and (\ref{eq: ohmic power-law noise spectrum}) (see Appendix \ref{sub:Dephasinh-Rate}
for details):

\noindent \emph{(i)} For \emph{generalized Ornstein-Uhlenbeck spectra
$S_{\beta}(\omega)$,} only high frequencies must be probed by the
dynamical control filter. To this end, the intervals between pulses
or refocusing periods must obey $\frac{t}{N}\ll\tau_{c}$.\textit{
}Then, $\mathcal{J}_{\beta}\propto\tau_{c}^{-(\beta-1)}$ satisfies
the equality in (\ref{eq:ultimate bound}) and the minimal relative
error attains the ultimate bound, provided that the total control
time is chosen to have the optimal value
\begin{equation}
t_{\beta}^{opt}=\tau_{c}\,\sqrt[\beta+1]{\frac{N^{\beta}\mathcal{J}_{0}}{c_{\beta}g^{2}\tau_{c}^{2}}},\label{eq:topt_b}
\end{equation}
where $c_{\beta}$ is a constant depending on the control sequence.
The bound can be attained only if $N$ is sufficiently large to satisfy
$t_{\beta}^{opt},\tau_{c}\gg\frac{t_{\beta}^{opt}}{N}$ and only overlaps
with the power-law tail, as shown in Fig. \ref{fig1_scheme}c. By
contrast, FID fails this condition, since its filter mainly overlaps
with $\omega\approx0$ (Fig. \ref{fig1_scheme}c), causing a larger
error in the estimation, as shown in Fig. \ref{fig2_opt-cntrl_Lorentz}b.

\noindent \emph{(ii)} For\emph{ generalized Ohmic spectra $S_{s}$},
the filter should only overlap with the noise spectrum at a frequency
lower than the cutoff, $0<\omega<\omega_{c}$, avoiding any overlap
at $\omega_{c}$. Then, $\mathcal{J}_{s}\propto\tau_{c}^{s+1}$ satisfies
the equality in (\ref{eq:ultimate bound}) and the minimal relative
error attains the ultimate-bound when the measurement of $\sigma_{x}$
is performed at the optimal total control time
\begin{equation}
t_{s}^{opt}=\tau_{c}\sqrt[s-1]{\frac{c_{s}g^{2}\tau_{c}^{2}N^{s}}{\mathcal{J}_{0}}}.\label{eq:topt_s}
\end{equation}
Here $\frac{t_{s}^{opt}}{N}>\tau_{c}$, and $N\gg1$ ensure the filter
is narrowband with negligible tail at $\omega_{c}$, as opposed to
its free-evolution (FID) counterpart that causes a much larger estimation
error (see Fig. \ref{fig2_opt-cntrl_Lorentz}b).

Equations (\ref{eq:topt_b}) and (\ref{eq:topt_s}) thus constitute
our main condition on the \emph{optimal filter design} for estimating
$\tau_{c}$.

\subsection{Power-law estimation\label{sub:Power-law-estimation}}

One can \emph{optimally estimate the exponent }$\beta$ or $s$ that
governs the bath spectrum by maximizing the QFI (\ref{eq:QFI_R})
for the estimation of $x_{B}=\gamma=\beta+1=1-s$ (Appendix \ref{sub:Dephasinh-Rate}).
This maximization leads to the ultimate precision bound
\begin{equation}
\varepsilon(\gamma,t)\!\geq\!\frac{1}{\gamma\sqrt{N_{m}\mathcal{F_{Q}}(\gamma,t)}}\!\geq\!\frac{\sqrt{1-e^{-2\mathcal{J}_{1}}}}{\sqrt{N_{m}}\,\mathcal{J}_{1}|ln(\mathcal{J}_{1})|\, e^{-\mathcal{J}_{1}}}\!=\!\frac{\varepsilon_{1}}{\sqrt{N_{m}}},
\end{equation}
with $\mathcal{J}_{1}=e^{-2.246}\approx0.106$ and $\varepsilon_{1}\approx2.04$.
The bound is achieved when the qubit only probes the power-law regime
of noise spectra (\ref{eq:power-law tail noise spectrum})-(\ref{eq: ohmic power-law noise spectrum}).
Provided that the CPMG or CW dynamical control probes the power law
region, the bound is attained when the qubit-probe is measured at
\begin{equation}
t_{opt}=T_{2}\sqrt[\gamma]{\mathcal{J}_{1}},\label{eq:topt-powerlaw}
\end{equation}
where $\mathcal{J}(\gamma,t)=\left(\frac{t}{T_{2}}\right)^{\gamma}$
and $T_{2}$ is the dephasing time (that depends on the applied control).

\section{\noindent \textit{\emph{Real-time estimation protocol}}}

\noindent \textbf{\textit{\emph{\label{sec:Real-time-estimation-protocol}}}}The
predicted optimal time $t^{opt}$ for performing a measurement following
the dynamical control or FID of the quantum-probe coherence, explicitly
depends on the \emph{unknown parameter} $x_{B}$ to be estimated (see
Eqs. (\ref{eq:toptZeno}), (\ref{eq:topt_b}), (\ref{eq:topt_s})
and (\ref{eq:topt-powerlaw})). In practice, one may bypass this difficulty
by \emph{estimating $x_{B}$ and simultaneously} \emph{finding the
optimal time} to monitor the probe via an efficient real-time estimation
protocol \cite{Cory2012_Robust-online-Hamiltonian-learning,Sergeevich_Characterization_2011,Cappellaro_2012_Spin-bath}.
We illustrate here the implementation of this protocol and show that
it attains the predicted ultimate bound (\ref{eq:ultimate bound})
if the qubit-probe is optimally controlled.

\noindent The protocol is based on a Bayesian estimator and an online
experimental learning design that maximizes the information gain,
similar to the one that was recently implemented for improving the
suppression of decoherence in quantum dots \cite{Yacoby2014}. Its
stages are as follows:

\noindent \emph{(i)} We initially guess a probability distribution
$p(x_{B})$ for the unknown parameter $x_{B}$ that represents our
apriori knowledge of the parameter to be estimated. Physically, we
should have $x_{B}>0$, and therefore one can assume a flat distribution
of $p(x_{B})$, $x_{B}\in(x_{B}^{min},x_{B}^{max}]$ that contains
the true value $x_{B}^{true}$. 

\noindent \emph{(ii)} We then determine the best time $t_{m}$ to
perform a measurement for maximizing \textit{the information gain}
about $x_{B}$ which is defined by the information entropy $U(t_{m})=\underset{t}{max}\left\{ U(t)\right\} $
(Appendix \ref{sub:Real-time-estimation-protocol}), which depends
on $p(x_{B})$ and the \textit{likelihood function} $p(d|x_{B},t)$
of Eq. (\ref{eq:p+-}) that determines the conditional probability
to obtain the possible outcome data of the measurement $d=\{+,-\}$.

\noindent \emph{(iii)} We next perform a measurement of $\sigma_{x}$
of the qubit-probe state, initialized in the state $\left|+\right\rangle $,
at the optimal time $t_{m}$, obtaining the outcome data $d$ \textit{\emph{with
probability}} $p(d|x_{B}^{true},t)$. According to the obtained data,
we update our knowledge of $x_{B}$ by the Bayesian Rule
\begin{equation}
p^{new}(x_{B})\equiv p(x_{B}|d,t_{m})=\frac{p(d|x_{B},t_{m})\, p(x_{B})}{p(d|t_{m})},\label{eq:BayesProb}
\end{equation}
where $p(d|t)$ is a normalization factor for integration over $x_{B}$.

\noindent The estimation of $x_{B}$ improves upon iteratively repeating
this three-stage process $N_{m}$ times, $N_{m}$ standing for the
number of measurements. When an adequate control is chosen, the probability
distribution $p(x_{B})$ converges to a narrow peak around $x_{B}^{true}$.
By contrast, the convergence under free-evolution can be very poor. 

\begin{figure*}
\includegraphics[width=0.7\textwidth]{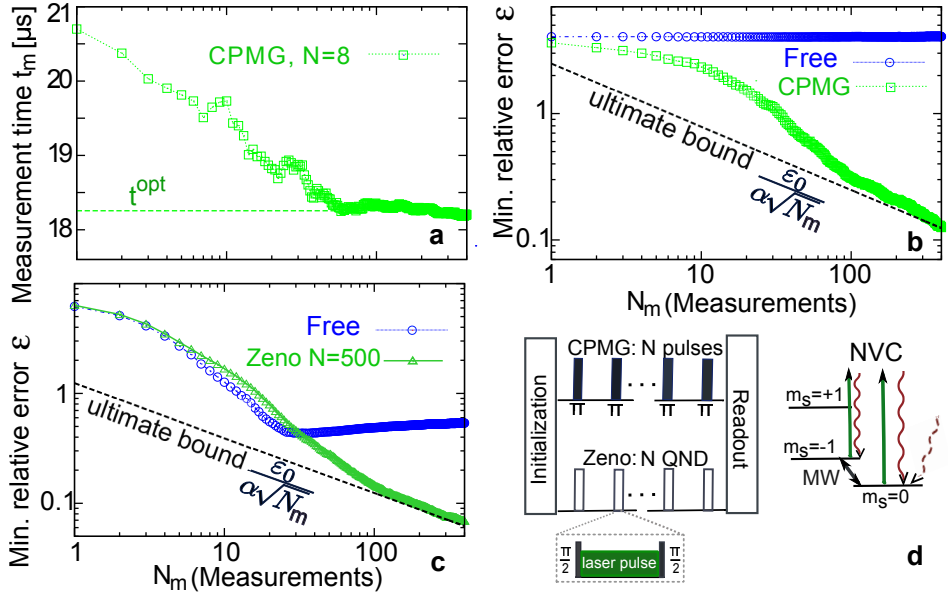}\protect\caption{\label{fig4_converg-algorith} Simulation of a experimental real-time
adaptive estimation protocol for realistic conditions with a NVC spin-probe.
(a,b) Convergence of the real-time adaptive estimation protocol to
the theoretically predicted values for estimating $\tau_{c}$. Free-evolution
of the probe (blue circles) is contrasted with that of a dynamically
controlled probe under CPMG (green square) sequence with $N=8$ in
the presence of an Ornstein-Uhlenbeck process with Lorentzian spectrum
$G_{\beta=2}(\tau_{c}^{true}\!\!=\!\!10\mu\mbox{s},\omega)$, with
$g=1$ MHz consistently with the spectral density of a HPHT diamond
sample determined in Ref. \cite{Bar-Gill2012}. The simulated curves
derived from exact analytical results of Eq. (\ref{eq:Dephasing rate xB})
\cite{Alvarez2013}, were averaged over 600 realizations. In (a) the
optimal measurement time $t_{m}$ as a function of $N_{m}$ converges
to the value $t^{opt}$ for the CPMG case. Similar curves converging
to the corresponding $t^{opt}$ are observed for other controls and
free evolution. In (b) the minimal relative error $\varepsilon(\tau_{c},t^{opt})$
converge to the (Cramer-Rao) bound. Under free evolution the regime
where $\varepsilon\propto\frac{1}{\sqrt{N_{m}}}$ is attained for
$N_{m}\gg100$. The \emph{ultimate bound }(\emph{$\frac{\varepsilon_{0}}{\sqrt{N_{m}}}$}
dashed line, $\alpha=\beta-1$)\emph{ is only attained by optimal
control}. (c) Convergence to the minimal relative error $\varepsilon(g,t_{opt})$
to the (Cramer-Rao) bound for estimating $g$ by $N=500$ consecutive
projective measurements in the Zeno regime (green triangle) compared
to the estimation under free evolution (blue circle). In this case
$G_{\beta=2}(g\!\!=\!\!0.03\mbox{MHz},\omega)$, with $\tau_{c}=10\mu s$,
consistently with the spectral density of a $^{\ensuremath{12}}$C
diamond sample determined in Ref. \cite{Bar-Gill2012}. Here too the
\emph{ultimate bound ($\frac{\varepsilon_{0}}{2\sqrt{N_{m}}}$} dashed
line, $\alpha=2$)\emph{ is only attained by optimal control}. (d)
Proposed scheme for using a NVC as a qubit-probe for its environment.
The $m_{s}=0$ ($\left|0\right\rangle $) state is fully populated
by laser irradiation (dashed curly arrow). Microwave (MW) pulses are
selectively applied between the states with $m_{s}=0$ and $-1$ ($\left|0\right\rangle $
and $\left|-1\right\rangle $) to initialize the spin-probe in a $\left|+\right\rangle =\frac{1}{\sqrt{2}}(\left|0\right\rangle +\left|-1\right\rangle )$
state, and to apply the $\pi$ pulse CPMG sequence for estimating
$\tau_{c}$. For estimating $g$, projective measurements are performed
by combining MW $\pi/2$ pulses on the $0\leftrightarrow-1$ transition
and laser-induced relaxation between the ground and exited electronic
states that conserve the spin components (solid curly arrows). The
readout is done at the end of the sequence by detecting the laser-induced
fluorescence signal.}
\end{figure*}

\noindent Figure \ref{fig4_converg-algorith} presents a simulated
experiment of this iterative process for the estimation of $g$ and
$\tau_{c}$ of the environmental noise by a NVC spin-probe in two
types of diamond samples whose intrinsic environmental characteristics
require dynamical control to attain the ultimate bound (\ref{eq:ultimate bound}):

\noindent (a) We simulate the estimation of $\tau_{c}$ for the NVC-probe
evolving freely (undergoing FID), as compared to it being nearly optimally
controlled by a CPMG sequence with $N=8$ subject to an Ornstein-Uhlenbeck
process characterized by $G_{\beta=2}(\tau_{c}^{true}\!\!=\!\!10\mu\mbox{s},\omega)$
with $g=1$ MHz (consistently with the HPHT diamond sample data of
Ref. \cite{Bar-Gill2012}). Our simulated best-measurement timings
$t_{m}$ converge to the theoretically predicted $t^{opt}$ that maximizes
the QFI (\ref{eq:QFI_R}), as exemplified in Fig. \ref{fig4_converg-algorith}a
for CPMG control, where $t^{opt}\approx18.25\mu s$ is on time scales
compatible with the accessible experimental times \cite{Bar-Gill2012}.
Concurrently, the minimum relative error converges to $\varepsilon(\tau_{c},t_{opt})$
predicted from the Cramer-Rao bound (Fig. \ref{fig4_converg-algorith}b).

\noindent (b) In Fig. \ref{fig4_converg-algorith}c we simulate the
estimation of $g$ under similar conditions by a sequence of $N=500$
QND measurements due to induced dephasings whose optimal intervals
are $\frac{t_{opt}}{N}\approx1.9\mu s$ and compare these results
to the FID results. In this case $G_{\beta=2}(g^{true}=0.03\mbox{MHz},\omega)$
with \foreignlanguage{english}{$\tau_{c}\!\!=\!\!10\mu\mbox{s}$ }is
consistent with the spectral density determined in Ref. \cite{Bar-Gill2012}
for the $^{\ensuremath{12}}$C diamond sample.
This estimation procedure is performed \emph{without assuming the
lineshape of the spectral density.} It is seen from Fig. \ref{fig4_converg-algorith}c
that the precision error eventually saturates for the free evolution
(FID) at a precision that is limited by our lack of knowledge of the
lineshape. By contrast, this saturation can be overcome in the QZE
regime achieved by projective measurements.

\noindent The foregoing simulations have thus confirmed, under experimentally
realistic conditions, our second major analytical result, whereby
the \emph{ultimate theoretical bound} (\ref{eq:ultimate bound}) is
\emph{indeed attainable under optimal control, but hardly ever under
free-evolution}.

\section{\noindent Discussion\label{sec:Discussion}}

\noindent We have demonstrated that dynamical control of a quantum
probe not only \emph{dramatically improves} the quantum estimation
of environmental parameters compared to the free-evolution (induction)
decay (FID) of its coherence: it may be \emph{imperative} to use such
control, since FID may preclude their correct estimation. In particular,
for generic noise spectra, as in generalized Ornstein-Uhlenbeck and
Ohmic processes, the ultimate analytical bounds for the coupling strength
and correlation-time estimation precision derived here can be achieved
by optimizing the dynamical control on the probe. The optimal controls
suitable for the estimation of $g$, $\tau_{c}$ (Fig. \ref{fig1_scheme},
\ref{fig4_converg-algorith}) or the power-law exponent are generally
different, but the protocol is similar. Specifically, we demonstrated
that the ultimate estimation bound for $g$ can be attained in the
quantum Zeno regime without prior knowledge of the environmental spectral-density
shape. Once this probe-environment coupling is known, the environmental
correlation time can be reliably estimated by standard dynamical-control
sequences. 

\noindent Experimental conditions typically require a minimized average
error within the coherence time interval shown in Fig. \ref{fig2_opt-cntrl_Lorentz}b,
if this interval is much shorter than the combined post-measurement
readout and initialization time of the qubit. Under such conditions,
an important implication of the present analysis is that it yields
the optimal \emph{total time} of the pre-measurement dynamical control
and thereby the \emph{number of pulses} and the delay between them
that need be applied to attain the best precision per measurement.
However, in the (rarely encountered) opposite limit of fast readout
and initialization, the overall error rate of the measurements is
to be minimized. Then, a similar optimization can be performed by
maximizing the Fisher information \emph{per unit time.} If additional
specific constraints are imposed by a particular experimental setup,
e.g. limited power or estimation time, then further optimization is
needed to find the optimal filter function. In such cases, the goal
would be to approach as best we can the ultimate error bound under
the given constraints.

A real-time adaptive estimation protocol has illustrated the ability
to find the optimal time of dynamical control for achieving the predicted
precision bounds, e.g., for a nitrogen vacancy center (NVC) in a diamond
that acts as a qubit-probe of the bulk or surface in optically detected
NMR and MRI. Its aim here is to determine the noise spectrum generated
by nuclear or electronic spins \cite{Mamin2013,Staudacher2013,Steinert2013,Grinolds2014,Sushkov2014},
as well as other sources of dephasing. In Fig. \ref{fig4_converg-algorith}
we illustrated this protocol for estimating both $g$ and $\tau_{c}$
near an NVC used as a probe. The inferred correlation times of the
noise fluctuations can be particularly helpful for studying molecular
diffusion at the nanoscale, where the power law tails of noise spectra
allow the extraction of the diffusion rates and restriction lengths
associated with pore structures and thereby characterizing biological
processes \cite{Alvarez2013,Alvarez2013_JMR}.

In other scenarios, power-law tails can also characterize charge diffusion
in conducting crystals \cite{Feintuch2004} or spin diffusion in complex
spin-networks \cite{Suter1985,Alvarez2011,alvarez_controlling_2012,Alvarez2013a}.
Generalized Ohmic spectra may help understanding the functioning of
nanoscale electromechanical devices \cite{Joachim2000}, as well as
superconducting devices attached to conducting leads \cite{Ithier2005}. 

Several extensions of the outlined strategy will be further explored: 

(a) If uncontrollable noise sources are present, such as an intrinsic
$T_{2}$ decoherence due to a white-noise (Markovian) process or due
to pulse imperfections, we may have to resort to more elaborate controls,
nicknamed by us ``selective dynamical recoupling'' (SDR), which
we have shown to allow \emph{selective} probing of the targeted noise
source \cite{Smith2012,Alvarez2013}. SDR was already implemented
to \emph{selectively} monitor diffusion processes characterized by
an Ornstein-Uhlenbeck spectral density, so as to determine the probe-environment
coupling strengths \cite{Smith2012,Alvarez2014} and the environmental
correlation time, as well as diffusion restriction lengths \cite{Alvarez2013,Alvarez2013_JMR}
in systems of biological and chemical interest. We envisage that by
incorporating SDR within the present optimized estimation strategy
one may eliminate Markovian (intrinsic-$T_{2}$ and pulse-error) effects
and allow a clean parameter estimation of the targeted noise spectrum. 

(b) A qubit-probe may be replaced by an entangled $n$-particle probe
that may yield a lower error bound \cite{Maccone_2006_Quantum,Maccone_2014_Using,Huelga_1997_Improvement}.
The estimation strategy will remain similar, but the dynamical control
will have to be adapted to a multipartite scenario, based on the approach
of Refs. \cite{gordon_Preventing_2006,clausen_bath-optimized_2010,gordon_scalability_2011}. 

(c) The present strategy, whereby environmental parameters are estimated
one by one, provides the ultimate precision bounds that can be attained
for estimating a single parameter, assuming that the remaining parameters
are known with much better accuracy. Therefore, these bounds are also
valid for more elaborate multiparameter estimation strategies based
on a quantum Fisher information matrix \cite{Caves_2011_Fundamental},
although these bounds may not be tight in the latter case. Still,
the single-parameter estimation expounded here may be superseded by
multi-parameter optimized estimation per measurement.

To conclude, we have analytically set the error bounds on environmental-parameter
estimation and demonstrated the ability to extract such environmental
information with maximal accuracy by the least number of measurements
possible upon invoking dynamical control. This general result for
a qubit-probe weakly coupled to its environment opens the door to
the development of an important diagnostic tool of environmental processes
by quantum probes.

\appendix

\section{\noindent Anamnesis: Filter function derivation for qubit probe under
dephasing (Following Refs {[}14-16,24-25{]})\label{sub:Quantum-probe-under}}

\noindent The Hamiltonian of a single spin-probe interacting with
a bath that produces pure dephasing is
\begin{equation}
H=H_{S}(t)+H_{B}+H_{SB},
\end{equation}
with
\begin{equation}
H_{S}=\frac{\omega_{0}}{2}\sigma_{z}+f(t)e^{-i\omega_{0}t}\sigma_{x},\: H_{SB}=S\otimes B=g\sigma_{z}B,
\end{equation}
where $f(t)$ is the dynamical control applied to the qubit, $B$
and $S$ are the operators of the bath and the system ($\sigma_{z}$
for pure dephasing) respectively, and $g$ is their coupling strength.

In the interaction picture the Hamiltonian can be written as 
\begin{equation}
H_{SB}^{I}(t)=S(t)\otimes B^{\dagger}(t),\label{eq:H^=00007BIntPictGral=00007D}
\end{equation}
 where 
\begin{equation}
\begin{array}{c}
S(t)=U_{S}^{\dagger}(t)SU_{S}(t),\, U_{S}(t)=\mathcal{T}e^{-i\intop_{0}^{t}dt^{'}H_{S}(t^{'})},\\
B(t)=U_{B}^{\dagger}(t)BU_{B}(t),\, U_{B}(t)=e^{-iH_{B}t}.
\end{array}
\end{equation}

\noindent Therefore, Eq. (\ref{eq:H^=00007BIntPictGral=00007D}) becomes
\begin{equation}
H_{SB}^{I}=g\Omega(t)\sigma_{z}B(t),
\end{equation}
where $\Omega(t)$ is the dynamical control rate of the system.

From this form one can derive the non-Markovian master equation for
the density matrix of the system, $\rho_{S}(t)$ in the interaction
picture, which in the Born approximation one assumes a weak coupling
$g$ such that the the influence of the qubit-probe on the environment
is small (usually called the weak-coupling approximation). In this
approximation, the density matrix of the environment $\rho_{B}$ is
only negligibly affected by the interaction with the qubit-probe,
and the state of the total system at time $t$ is allowed to be expressed
as $\rho(t)\approx\rho_{S}(t)\otimes\rho_{B}$ \cite{Breuer2007}.
The resulting non-Markovian master equation is then given by \cite{kofman_universal_2001,kofman_unified_2004,Kofman_Theory_2005,gordon_universal_2007,gordon_dynamical_2009}
\begin{equation}
\dot{\rho}_{S}(\tau_{c},t)=\intop_{0}^{t}dt'\{g^{2}\Phi(x_{B},t-t')[S(t'),S(t)\rho_{S}(t)]+h.c\},
\end{equation}
where $\Phi(x_{B},t'-t")=\mathrm{Tr}_{B}\left\{ B(t'-t")B(0)\rho_{B}(0)\right\} $
are the bath correlation functions and $x_{B}$ is a parameter that
characterize the environment. Then, the attenuation factor of the
spin coherence (magnetization) $\langle\sigma_{x}(t)\rangle=e^{-\mathcal{J}(x_{B},t)}$
measured at time $t$ is
\begin{equation}
\mathcal{J}(x_{B},t)=\intop_{0}^{t}dt'\intop_{0}^{t'}dt"g^{2}\Phi(x_{B},t'-t")\Omega(t')\Omega^{*}(t"),
\end{equation}
which can be cast in the spectral form
\begin{equation}
\mathcal{J}(x_{B},t)=\int_{-\infty}^{\infty}d\omega F_{t}(\omega)G(x_{B},\omega),\label{eq:Dephasing rate}
\end{equation}
\begin{equation}
G(x_{B},\omega)=\frac{1}{2\pi}\int_{-\infty}^{\infty}dtg^{2}\Phi(x_{B},t)e^{i\omega t},
\end{equation}
 being the bath-coupling spectrum and
\begin{equation}
F_{t}(\omega)=\frac{1}{2\pi}\left|\int_{0}^{t}dt'\Omega(t')e^{i\omega t'\text{ }}\right|^{2}
\end{equation}
 is the filter function which depends on the dynamical control of
the qubit probe.

We stress that $\Omega(t)$ can have an \emph{arbitrary} temporal
shape, as opposed to the restrictions on its (\emph{inherently pulsed})
shape in the dynamical decoupling method (Refs. {[}12,13,17,27{]}
in the main text).

\section{\noindent Quantum Fisher information concerning a single parameter
of the environment\label{sub:Quantum-Fisher-information} }

\noindent The quantum Fisher information (QFI) concerning a single
parameter $x_{B}$ of the environment (bath) for the qubit state is
\cite{Caves_1994_fisher,Paris2009_QUANTUM-ESTIMATION,Paris2014_Characterization-of-classical-Gaussian}

\begin{align}
\mathcal{F_{Q}=} & \frac{1}{p_{+}}\left(\frac{\partial p_{+}}{\partial x_{B}}\right)^{2}+\frac{1}{p_{-}}\left(\frac{\partial p_{-}}{\partial x_{B}}\right)^{2}\label{eq:QFI-x_B}\\
 & +2\frac{(p_{+}-p_{-})^{2}}{p_{+}+p_{-}}\left(\left|\left\langle p_{-}\right|\frac{\partial\left|p_{+}\right\rangle }{\partial x_{B}}\right|+\left|\left\langle p_{+}\right|\frac{\partial\left|p_{-}\right\rangle }{\partial x_{B}}\right|^{2}\right),\nonumber 
\end{align}
where
\begin{align}
p_{\pm}(x_{B}t)\equiv & p(\pm|x_{B},t)=\frac{1}{2}\left(1\pm e^{-\mathcal{J}(x_{B},t)}\right),\\
\left|p_{\pm}\right\rangle = & \frac{1}{\sqrt{2}}\left(e^{-i\omega_{0}t}\left|\uparrow\right\rangle \pm\left|\downarrow\right\rangle \right)\nonumber 
\end{align}
are the eigenvalues and eigenvectors of the spin-probe density matrix
\cite{gordon_universal_2007,gordon_dynamical_2009}. A measurement
is said to be optimal when the QFI $\mathcal{F_{Q}}$ coincides with
its classical counterpart \cite{Caves_1994_fisher,Paris2009_QUANTUM-ESTIMATION,Paris2014_Characterization-of-classical-Gaussian}.
This is the case here under pure dephasing, when the last term in
(\ref{eq:QFI-x_B}) is null due to $\frac{\partial\left|p_{+}\right\rangle }{\partial x_{B}}=0$.
Then the optimal measurement is effected by projections onto the eigenstates
$\left|p_{\pm}\right\rangle $ of $\sigma_{x}$ in the rotating frame
\begin{align}
\left|p_{\pm}\right\rangle \left\langle p_{\pm}\right|= & \frac{1}{2}e^{\text{\textminus}i\frac{\omega_{0}}{2}\sigma_{z}}\left|\pm\right\rangle \left\langle \pm\right|e^{i\frac{\omega_{0}}{2}\sigma_{z}},\\
\left|\pm\right\rangle = & \frac{1}{\sqrt{2}}\left(\left|\uparrow\right\rangle \pm\left|\downarrow\right\rangle \right).\nonumber 
\end{align}

Correspondingly, Eq. (\ref{eq:QFI-x_B}) becomes 
\begin{equation}
\mathcal{F_{Q}}(x_{B},t)=\frac{e^{-2\mathcal{J}(x_{B},t)}}{1-e^{-2\mathcal{J}(x_{B},t)}}\left(\frac{\partial\mathcal{J}(x_{B},t)}{\partial x_{B}}\right)^{2},\label{eq:QFI-J}
\end{equation}
if the initial probe-state is $\left|+\right\rangle $. An arbitrary
initial state, $\left(\cos(\theta)\left|\uparrow\right\rangle +i\,\sin(\theta)\left|\downarrow\right\rangle \right)$,
$0<\theta<\frac{\pi}{2}$, leads to $\mathcal{F_{Q}}(x_{B},t)\propto\sin^{2}(2\theta)$
\cite{Paris2014_Characterization-of-classical-Gaussian}. Therefore,
the optimal initial state leading to the maximal QFI is obtained for
$\theta=\frac{\pi}{4}$, \foreignlanguage{english}{$\left|+\right\rangle =\frac{1}{\sqrt{2}}\left(\left|\uparrow\right\rangle +\left|\downarrow\right\rangle \right),$
thus proving Eq. (6) in the main text.}

\section{\noindent Derivation of the ultimate precision bound\label{sub:Bound-derivation}}

A \textit{\emph{broad class of practically relevant environmental
}}noise processes\textit{\emph{ }}cause the attenuation factor to
have a power-law functional dependence in $x_{B}$ with exponent $\alpha$,
where the derivative of $\mathcal{J}(x_{B},t)$ with respect to $x_{B}$
satisfies 
\begin{equation}
\left|\frac{\partial\mathcal{J}(x_{B},t)}{\partial x_{B}}\right|\le\frac{\alpha\mathcal{J}(x_{B},t)}{x_{B}}.\label{eq: deriv-Att-fact_x_B-1}
\end{equation}
 The QFI then conforms to the inequality
\begin{equation}
\mathcal{F_{Q}}(x_{B},t)\le\frac{e^{-2\mathcal{J}(x_{B},t)}}{1-e^{-2\mathcal{J}(x_{B},t)}}\frac{\alpha^{2}\mathcal{J}(x_{B},t)}{x_{B}^{2}}^{2}.\label{eq:QFI-J-x_B}
\end{equation}

The equality is obtained when the attenuation factor is an homogeneous
function of degree $\alpha$, i.e. strictly obeys a power law. The
maximum of the QFI, regardless of the kind of control applied, is
\foreignlanguage{english}{then obtained when $\left|\frac{\partial\mathcal{J}(x_{B},t)}{\partial x_{B}}\right|=\frac{\alpha\mathcal{J}(x_{B},t)}{x_{B}}$
and $\mathcal{J}(x_{B},t)=\mathcal{J}_{0}=1+\frac{1}{2}W(-2e^{-2})\approx0.8$},
where $W(z)$ is the Lambert function which by definition satisfies
$z=W(z)e^{W(z)}$ for any complex number $z$. When the control is
such that the equality in (\ref{eq:QFI-J-x_B}) is satisfied, then
the optimal total control-time at which the measurement should be
done, $t_{opt}$, is such that $\mathcal{J}(x_{B},t_{opt})=\mathcal{J}_{0}$. 

Under this condition, the resulting ultimate bound for the relative
error in the estimation of $x_{B}$, which holds for arbitrary control
on the probe, is

\noindent 
\begin{equation}
\varepsilon(x_{B},t)\!\geq\!\frac{1}{x_{B}\sqrt{N_{m}\mathcal{F_{Q}}(x_{B},t)}}\!\geq\!\frac{\sqrt{1-e^{-2\mathcal{J}_{0}}}}{\alpha\sqrt{N_{m}}\,\mathcal{J}_{0}\, e^{-\mathcal{J}_{0}}}\!=\!\frac{\varepsilon_{0}}{\alpha\sqrt{N_{m}}},\label{eq:Relative-Error-x_B}
\end{equation}
with $\varepsilon_{0}=\sqrt{\frac{2}{-W(-2e^{-2})(1+\frac{1}{2}W(-2e^{-2}))}}\approx2.48$.

\subsection{\noindent Precision bounds for the key parameters $g$ and $\tau_{c}$}

\subsubsection{\noindent (a) The probe-bath interaction strength $g$}

The attenuation factor is an homogeneous function of degree $\alpha=2$
in the effective probe-bath interaction strength $g$, where $G(\tau_{c},\omega)=g^{2}S(\omega)$
with \foreignlanguage{english}{$S(\omega)$} the normalized spectral
density of the environmental noise. Therefore, Eqs. (\ref{eq: deriv-Att-fact_x_B-1})
and (\ref{eq:QFI-J-x_B}) are satisfied and the minimal relative error
in the estimation of $g$ is obtained by measuring at the time $t_{opt}$,
such that $\mathcal{J}(g,t_{opt})=\mathcal{J}_{0}$, if \foreignlanguage{english}{$S(\omega)$
is known}.

\subsubsection{\noindent (b) The correlation time $\tau_{c}$}

\noindent The derivative term in the QFI of Eq. (\ref{eq:QFI-J})
depends on\foreignlanguage{english}{ the derivative of the bath coupling-spectrum
(spectral density) with respect to the correlation time $\frac{\partial G}{\partial\tau_{c}}$,
through}

\begin{equation}
\frac{\partial\mathcal{J}(\tau_{c},t)}{\partial\tau_{c}}=\int_{-\infty}^{\infty}d\omega F_{t}(\omega)\frac{\partial G(\tau_{c},\omega)}{\partial\tau_{c}}.\label{eq:dJ}
\end{equation}
Spectral densities of the baths characterized by $G_{\beta(s)}(\tau_{c},\omega)=g^{2}S(\omega)$
{[}defined in the main text, Eqs. (7-8){]} satisfy $\left|\frac{\partial S_{\beta(s)}}{\partial\tau_{c}}\right|\le\frac{\alpha S_{\beta(s)}}{\tau_{c}}$.
Specifically,
\begin{align}
\left|\frac{\partial G_{\beta}}{\partial\tau_{c}}\right| & =\left|\frac{\partial}{\partial\tau_{c}}\left(\frac{\mathcal{A}_{\beta}g^{2}\tau_{c}}{1+\omega^{\beta}\tau_{c}^{\beta}}\right)\right|\label{eq:derGb}\\
 & =\left|\frac{\mathcal{A}_{\beta}g^{2}}{1+\tau_{c}^{\beta}\omega^{\beta}}\left(1-\frac{\beta\tau_{c}^{\beta}\omega^{\beta}}{1+\tau_{c}^{\beta}\omega^{\beta}}\right)\right|\nonumber \\
 & \leq\frac{\mathcal{A}_{\beta}g^{2}(\beta-1)}{(1+\tau_{c}^{\beta}\omega^{\beta})}=\frac{(\beta-1)G_{\beta}}{\tau_{c}}\nonumber 
\end{align}
where the equality is attained when the spectral density function
behaves as an homogeneous function of degree $\alpha=\beta-1$, $G_{\beta}\propto\tau_{c}^{-(\beta-1)}\omega^{\beta}$;
and 

\selectlanguage{english}%
\begin{equation}
\frac{\partial G_{s}}{\partial\tau_{c}}=\frac{\partial}{\partial\tau_{c}}\left((s+1)\tau_{c}^{s+1}\omega^{s}\right)=\frac{(s+1)}{\tau_{c}}G_{s}\label{eq:derGs}
\end{equation}
\foreignlanguage{american}{since $G_{s}$ is an homogeneous function
of degree $\alpha=s+1$ upon neglecting the cutoff region.}

\selectlanguage{american}%
Using Eqs. (\ref{eq:derGb}-\ref{eq:derGs}) to bound Eq. (\ref{eq:dJ}),
we find
\begin{align}
\left|\frac{\partial\mathcal{J}(\tau_{c},t)}{\partial\tau_{c}}\right| & =\int_{-\infty}^{\infty}d\omega F_{t}(\omega)\left|\frac{\partial G(\tau_{c},\omega)}{\partial\tau_{c}}\right|\\
 & \leq\frac{\alpha}{\tau_{c}}\int_{-\infty}^{\infty}d\omega F_{t}(\omega)G(\tau_{c},\omega)\nonumber \\
 & =\frac{\alpha\mathcal{J}(\tau_{c},t)}{\tau_{c}}.\nonumber 
\end{align}
This leads to the tight bound for the Fisher information (\ref{eq:QFI-J-x_B})
and therefore, for the minimal error in the estimation (\ref{eq:Relative-Error-x_B})
which holds for arbitrary control on the spin-probe. The ultimate
bound in the precision (\ref{eq:Relative-Error-x_B}) is attained
when the optimal total control-time at which the measurement should
be done, $t_{opt}$, is such that $\mathcal{J}(\tau_{c},t_{opt})=\mathcal{J}_{0}$.

\section{\noindent Attainment of the ultimate bound\label{sub:Dephasinh-Rate}}

\subsection{\noindent Attainment of the ultimate precision bound under optimal
dynamical control in the estimation of key parameters:}

\subsubsection{\noindent (i) Probe-bath interaction strength $g$}

The attenuation factor is an homogeneous function of degree $\alpha=2$
in the probe-bath interaction strength $g$. Therefore, Eqs. (\ref{eq: deriv-Att-fact_x_B-1})
and (\ref{eq:QFI-J-x_B}) are satisfied and the minimal relative error
in the estimation of $g$ is attained by measuring at a time $t_{opt}$
such that $\mathcal{J}(g,t_{opt})=\mathcal{J}_{0}$ if \foreignlanguage{english}{$S(\omega)$}
is known. 

If \foreignlanguage{english}{$S(\omega)$} is not known, then some
constraints apply for \foreignlanguage{english}{$t_{opt}$}. Considering
that the attenuation factor is given by (\ref{eq:Dephasing rate}),
then, if the the filter function $F_{t}(\omega)$ is much wider than
$G(g,\omega)=g^{2}S(\omega)$, it can be considered as a constant
$F_{t}$ in the integral (\ref{eq:Dephasing rate}). The attenuation
factor following the integration is then $g^{2}F_{t}$, where we have
used the normalization property of the spectral density $S(\omega)$.
This limit holds when interval is shorter than the correlation time
$\tau_{c}$ of the environment, so that the qubit-probe evolves freely,
yielding \cite{kofman_universal_2001,kofman_unified_2004,Kofman_Theory_2005,gordon_universal_2007,gordon_dynamical_2009}
\begin{equation}
F_{t}^{free}(\omega)=\frac{t^{2}}{2}\textrm{sinc\ensuremath{^{2}(\omega t)}}\approx\frac{t^{2}}{2}\textrm{ if }t\ll\tau_{c},\label{eq:Ft^free}
\end{equation}
and
\begin{equation}
\mathcal{J}^{free}(g,t)\approx\frac{1}{2}g^{2}t^{2}.\label{eq:J^free}
\end{equation}
Then, the bound in the estimation error (\ref{eq:Relative-Error-x_B})
is achieved when 
\begin{equation}
t_{opt}^{free}=\frac{\sqrt{2\mathcal{J}_{0}}}{g},\mathrm{\textrm{ provided }}t_{opt}^{free}=\sqrt{\frac{2\mathcal{J}_{0}}{g^{2}}}\ll\tau_{c}.\label{eq:estimt g, t_opt^free}
\end{equation}
This condition can only be fulfilled for large enough $\tau_{c}$.
To overcome this limitation, we may exploit dynamical control by means
of frequent, stroboscopic quantum non-demolition (QND) measurements
of the qubit-probe. If $N$ QND \emph{unread} measurements are performed
during a total time $t$, the dynamics conforms to Zeno regime with
an attenuation factor \cite{kofman_universal_2001,kofman_unified_2004,Kofman_Theory_2005,gordon_universal_2007,gordon_dynamical_2009}
\begin{equation}
\mathcal{J}^{Zeno}(g,t)=\frac{g^{2}t^{2}}{2N}\textrm{ if }\frac{t}{N}\ll\tau_{c}.
\end{equation}
Then, the condition for attaining the bound in the estimation error
(\ref{eq:Relative-Error-x_B}) is relaxed by $1/\sqrt{N}$, as
\begin{equation}
t_{opt}^{Zeno}=\frac{\sqrt{2N\mathcal{J}_{0}}}{g},\mathrm{\textrm{ provided }}\frac{t_{opt}^{Zeno}}{N}=\sqrt{\frac{2\mathcal{J}_{0}}{g^{2}N}}\ll\tau_{c}.
\end{equation}
The latter condition can be attained for
\begin{equation}
N\gg\frac{2\mathcal{J}_{0}}{g^{2}\tau_{c}^{2}}.
\end{equation}
In order to ensure that (\ref{eq:Dephasing rate}) is satisfied, suffice
it that the product $g\tau_{c}$ be \emph{roughly estimated,} by observing
\emph{the change in the decay law} from the anti-Zeno (AZE) or Fermi
Golden Rule to the QZE regime as $N$ increases \cite{Durga_Zeno_2011}.

\subsubsection{\noindent (ii) Correlation time $\tau_{c}$}

\noindent Control of the spin-probe by $N$-pulse CPMG sequences or
$N$-cycles of continuous-wave (CW) driving, leads to a filter function
\foreignlanguage{english}{$F_{t}(\omega)$} that converges to a Fourier
series described by a sum of delta functions (band narrow filters)
centered at the harmonics of the inverse cycle time, \foreignlanguage{english}{$k\omega_{0}=\frac{\pi kN}{t}$
$k\in\mathbb{N}$}, provided the total control time exceeds the interval
between the pulses $t\gg\frac{t}{N}$. Under these conditions, the
attenuation factor (\ref{eq:Dephasing rate}) becomes \cite{Alvarez2011,Ajoy2011}

\begin{equation}
\mathcal{J}(\tau_{c},t)=g^{2}\overset{\infty}{\underset{k=1}{\sum}}F_{t}(k\omega_{0})G(\tau_{c},k\omega_{0}).\label{eq:Attenuation-factor-J_N>>1}
\end{equation}

In what follows we infer the conditions for attaining the bound from
the attenuation factor for the bath spectra considered in the main
text.

\emph{(1)} For \emph{generalized Ornstein-Uhlenbeck spectra} $G_{\beta}$,
one condition to achieve the bound of Eq. (\ref{eq:Relative-Error-x_B})
is that the filter only overlap with the power-law tail spectra. This
is ensured when the first harmonic of the filter is already in this
spectral region, which amounts to $\frac{t}{\pi N}\ll\tau_{c}$. The
attenuation factor (\ref{eq:Attenuation-factor-J_N>>1}) then becomes
\begin{equation}
\mathcal{J}_{\beta}(\tau_{c},t)=g^{2}\overset{\infty}{\underset{k=1}{\sum}}F_{t}\left(\frac{\pi kN\tau_{c}}{t}\right)\frac{\mathcal{A_{\beta}}\tau_{c}}{\left(\frac{\pi kN\tau_{c}}{t}\right)^{\beta}}=\frac{c_{\beta}g^{2}t^{\beta+1}}{N^{\beta}\tau_{c}^{\beta-1}},\label{eq:Jb}
\end{equation}
with $\mathcal{A_{\beta}}=\frac{\beta}{2\pi}sin(\frac{\pi}{\beta})$
and $c_{\beta}=\frac{\beta}{2\pi^{\beta}}sin(\frac{\pi}{\beta})$
for CW (only the first harmonic $k=1$ gives non-null terms) and $c_{\beta}=\frac{\zeta(\beta+2)(4-2^{-\beta})\beta}{\pi^{2\beta}}sin(\frac{\pi}{\beta})$
for CPMG (only odd $k$ gives non-null terms) where $\zeta$ is the
zeta function defined for $Re(z)>1$ as $\zeta(z)=\Sigma_{i=1}^{\infty}\frac{1}{i^{z}}$.
Both constants have similar values $c_{\beta}^{CW}\approx c_{\beta}^{CPMG}$.

Since $\mathcal{J}_{\beta}$ satisfies the equality in Eq. (\ref{eq:QFI-J-x_B}),
the ultimate bound (\ref{eq:Relative-Error-x_B}) is achieved when
$\mathcal{J_{\beta}}(\tau_{c},t_{opt})=\mathcal{J}_{0}$. This yields
\begin{equation}
t_{\beta}^{opt}=\tau_{c}\,\sqrt[\beta+1]{\frac{N^{\beta}\mathcal{J}_{0}}{c_{\beta}g^{2}\tau_{c}^{2}}.}
\end{equation}
The resulting requirement on $N$, in order to keep the optimal narrowband
approximation and overlap only with the power law tail
\begin{equation}
\frac{t^{opt}}{\pi N}\ll\tau_{c},\frac{t^{opt}}{\pi}
\end{equation}
 is then
\begin{equation}
N\gg max\left\{ \frac{\mathcal{J}_{0}}{c_{\beta}g^{2}\tau_{c}^{2}\pi^{\beta+1}},1\right\} .
\end{equation}

If the control is constrained (say, by maximum total energy \cite{gordon_optimal_2008,clausen_bath-optimized_2010,clausen_task-optimized_2012}),
it may happen that the filter-function overlaps only with the Markovian
region of the spectral density, $G_{\beta}^{M}(\tau_{c},\omega)\approx\mathcal{A}_{\beta}\tau_{c}$,
where it becomes a homogeneous function of order $\alpha=1$ on $\tau_{c}$.
The attenuation factor is then independent of the dynamical control
and is given by the Markovian limit
\begin{equation}
\mathcal{J}^{M}(\tau_{c},t)=g^{2}\tau_{c}t.
\end{equation}
The ultimate bound (with $\alpha=1$) will then be achieved if the
probe-state is measured at $t_{\beta}^{M,opt}=\frac{\mathcal{J}_{0}}{g^{2}\tau_{c}}$
independently of the control. This bound is always greater (worse)
than the ultimate bound of Eq. (\ref{eq:Relative-Error-x_B}) for
$\beta>2$. For $\beta=2$, where they are equal.

\emph{(2) }For\emph{ generalized Ohmic spectra} $G_{s}$, the attenuation
factor (\ref{eq:Attenuation-factor-J_N>>1}) is
\begin{align}
\mathcal{J}_{s}(\tau_{c},t) & =g^{2}\overset{k_{c}}{\underset{k=1}{\sum}}F_{t}\left(\frac{\pi kN\tau_{c}}{t}\right)(s+1)\tau_{c}{}^{(s+1)}\left(\frac{\pi kN\tau_{c}}{t}\right)^{s}\\
 & =\frac{c_{s}g^{2}\tau_{c}^{s+1}N^{s}}{t^{s-1}},\nonumber 
\end{align}
where $c_{s}=\frac{\pi^{s+1}(s+1)}{2}\sum_{k=1}^{k_{c}}(2k-1)^{s-2}$
with $k_{c}=\left[\frac{t}{\pi N\tau_{c}}\right]$ where the square
brackets denote the integer part. Since the harmonics that contribute
with non-null terms are those below the cutoff, then $k_{c}$ is defined
so as to satisfy $\frac{\pi k_{c}N}{t}<\omega_{c}<\frac{\pi(k_{c}+1)N}{t}$
and therefore $\frac{t}{\pi N\tau_{c}}+1<k_{c}<\frac{t}{\pi N\tau_{c}}$. 

The first requirement to achieve the bound (\ref{eq:Relative-Error-x_B})
is to avoid any overlap between the filter and the cutoff region,
i.e. $\frac{t}{\pi Nk}\neq\tau_{c}$, with $k\in\mathbb{N}$. This
is most likely to be satisfied under the narrowband approximation
that allows to design the filter to be null at cutoff. 

The second requirement is that the filter should overlap with the
power-law region. Therefore, one needs at least one harmonic below
the cutoff that avoids the overlap with the cutoff, i.e.
\begin{equation}
\frac{t}{\pi N}>\tau_{c}.
\end{equation}
Both conditions are ensured under CW control if $\frac{t}{\pi N}>\tau_{c}$,
since the corresponding filter contains only one harmonic, leading
to a simplified expression for the constant $c_{s}=\frac{\pi^{s+1}(s+1)}{2}$.
Then, $\mathcal{J}_{s}$ satisfies the equality in Eq. (\ref{eq:QFI-J-x_B})
and the ultimate bound (\ref{eq:Relative-Error-x_B}) is achieved
when $\mathcal{J}_{s}(\tau_{c},t_{opt})=\mathcal{J}_{0}$, yielding
\begin{equation}
t_{s}^{opt}=\tau_{c}\sqrt[s-1]{\frac{c_{s}g^{2}\tau_{c}^{2}N^{s}}{\mathcal{J}_{0}}}.
\end{equation}
To maintain this narrowband approximation, with one harmonic below
the cutoff frequency, \emph{i.e.}
\begin{equation}
\frac{t_{s}^{opt}}{\pi}\gg\frac{t_{s}^{opt}}{\pi N}>\tau_{c},
\end{equation}
the number of cycles $N$ for super-Ohmic spectra ($s>1)$ should
satisfy
\begin{equation}
N_{superOhm}>\frac{\mathcal{J}_{0}\pi^{s-1}}{c_{s}g^{2}\tau_{c}^{2}}\textrm{ and }N_{superOhm}\gg1
\end{equation}
and for sub-Ohmic spectra ($0<s<1)$
\begin{equation}
1\ll N_{subOhm}<\frac{\mathcal{J}_{0}\pi^{s-1}}{c_{s}g^{2}\tau_{c}^{2}}
\end{equation}
which is attainable when $\sqrt{\frac{\mathcal{J}_{0}\pi^{s-1}}{c_{s}}}\gg g\tau_{c}$.
For Ohmic spectra ($s=1$) 
\begin{equation}
1\ll N_{Ohm}=\frac{\mathcal{J}_{0}}{c_{s}g^{2}\tau_{c}^{2}}
\end{equation}
implying that it must satisfy $\sqrt{\frac{\mathcal{J}_{0}}{c_{s}}}\gg g\tau_{c}$.

When the last conditions cannot be satisfied, less restrictive optimal
solutions can be found by demanding the filter function to be null
at the cutoff frequency $\omega_{c}$, and then finding the optimal
control time.

\subsubsection{\noindent (b) The power law exponent}

To estimate the power law (PL) exponents $s$ and $\beta$ of spectral
densities: $G^{PL}(\omega)\propto\omega^{s},\omega^{-\beta}$ or $G(\omega)\propto\frac{1}{1+a\omega^{\beta}}$
with $a=\mbox{const.}$, we define the exponent $x_{B}=\gamma$ that
is common to both spectral densities: $\gamma=1+\beta$ or $\gamma=1-s$,
we note that the derivative of the attenuation factor with respect
to the exponent satisfies (cf. Eqs. (1), (4) in the main text)
\begin{align}
\left|\frac{\partial\mathcal{J}(x_{B}=\gamma,t)}{\partial\gamma}\right| & =\left|\int_{-\infty}^{\infty}d\omega F_{t}(\omega)\frac{\partial G(\gamma,\omega)}{\partial\gamma}\right|\\
 & \le\left|\int_{-\infty}^{\infty}d\omega F_{t}(\omega)\frac{\partial G^{PL}(\gamma,\omega)}{\partial\gamma}\right|\nonumber \\
 & =\left|\frac{\partial\mathcal{J}^{PL}(\gamma,t)}{\partial\gamma}\right|.\nonumber 
\end{align}
This expression tightly bound the QFI (\ref{eq:QFI-J}).To attain
the ultimate bound for the relative error of the estimation by the
maximized QFI, one needs to apply a control that only probes the power-law
regime of the spectral density. The attenuation factor under such
control can be expressed as $\mathcal{J}(\gamma,t)=\mathcal{J}^{PL}(\gamma,t)=\left(\frac{t}{T_{2}}\right)^{\gamma}$,
where the dephasing time $T_{2}$ depends on the applied control.
Then, 

\begin{equation}
\left|\frac{\partial\mathcal{J}(\gamma,t)}{\partial\gamma}\right|=\frac{\mathcal{J}(\gamma,t)}{\gamma}|ln(\mathcal{J}(\gamma,t))|
\end{equation}
and the ultimate precision bound in the estimation is then given by
Eq. (13) of the main text. The bound is achieved when the qubit-probe
is measured at $t_{opt}=T_{2}\sqrt[\gamma]{\mathcal{J}_{1}}$ with
$\mathcal{J}_{1}=e^{-2.246}\approx0.106$ when the dynamical control
can ensure this regime. For example, under CW or CPMG control for
estimating the power law of an Ornstein-Uhlenbeck process, the time
interval between the number of pulses should be smaller than the correlation
time, $\frac{t_{opt}}{N}\ll\tau_{c}$, so that $N\gg\frac{T_{2}\sqrt[\gamma]{\mathcal{J}_{1}}}{\tau_{c}}$.

\subsubsection{\noindent (c) Unattainability of the ultimate bound under free evolution
(free-induction decay - FID)}

We here discuss the unattainability of the ultimate precision bound
when the spin-probe evolves freely, for the estimation of: 

(i) \emph{probe-bath interaction strength}\textbf{\emph{ }}\emph{$g$}
(the conditions to attain the relevant bound are discussed in the
main text starting with Eqs. (\ref{eq:Ft^free}) and ending in Eqs.
(\ref{eq:estimt g, t_opt^free})), and

(ii) \emph{the correlation time $\tau_{c}$}, for which, as observed
in Fig. 2 of the main text, the estimation precision becomes worse
as $g\tau_{c}$ grows. For the generalized Ornstein-Uhlenbeck spectra
$G_{\beta}$, as discussed in the main text, the free-evolution filter
$F_{t}^{free}(\omega)$, Eq. (\ref{eq:Ft^free}), overlaps with the
$\omega\approx0$ region and consequently the equality in Eq. (\ref{eq:derGb})
cannot be fulfilled. For the generalized Ohmic spectra $G_{s}$, when
$t\ll\tau_{c}$, one can approximate Eq. (\ref{eq:Dephasing rate})
by its zero-order term, considering that \foreignlanguage{english}{$F_{t}^{free}(\omega)$}
is independent of $\omega$, since \foreignlanguage{english}{$G(\tau_{c},\omega)$}
is much narrower than \foreignlanguage{english}{$F_{t}^{free}(\omega)$}. 

The attenuation factor in this regime becomes independent of $\tau_{c}$,
\foreignlanguage{english}{as in Eq. }(\ref{eq:J^free}), $\mathcal{J}_{\beta}^{free}(\tau_{c},t)=\mathcal{J}_{s}^{free}(\tau_{c},t)\approx\frac{1}{2}g^{2}t^{2}$.\foreignlanguage{english}{
}In this regime there is no information concerning the correlation
time $\tau_{c}$. If $\frac{1}{2}g^{2}\tau_{c}^{2}\gg\mathcal{J}_{0}$,
then it will not be possible to achieve the bound under free evolution.
The attenuation factor will depend on $\tau_{c}$ only when $t\gtrsim\tau_{c}$,
and therefore $\frac{1}{2}g^{2}t^{2}\gtrsim\frac{1}{2}g^{2}\tau_{c}^{2}\gg\mathcal{J}_{0}$,
implying that the ultimate bound on Eq. (\ref{eq:Relative-Error-x_B})
cannot be achieved. 

The limitation of the estimation of $\tau_{c}$ for the spectra $G_{s}$
using freely evolving spin-probes is that the only control parameter
is the time $t$ that may not simultaneously avoid the overlap of
the filter function with $\omega_{c}$ and render the attenuation
factor equal to $\mathcal{J}_{0}$, \emph{i.e.} usually $t_{opt}\neq4\pi n\tau_{c}$,
$n\in\mathbb{N}$. Hence, the ultimate bound is generally not achieved
under free evolution.

\subsubsection{\noindent (d) Pulse-error effects}

In the preceding analysis, we have considered ideal, i.e. perfect
and stroboscopic, pulses. If non-ideal pulse-effects are important,
one needs a model for their effect on the dephasing. Typically the
pulses are applied with the same phase, as in CPMG, where flip-angle
errors are compensated and the pure dephasing assumption is still
suitable \cite{Alvarez2010a,Ajoy2011}. In this case, if finite-width
pulse effects are significant, they can be considered by modifying
the effective filter functions \cite{Biercuk2009}. Therefore, within
this model for the error effects, our ultimate bound is valid and
achievable in the presence of such pulse-error effects. If pulse imperfections
generate an effective Hamiltonian that deviates from a pure dephasing
\cite{Alvarez2010a,Sli07}, then a different approach to calculate
the ultimate bound must be pursued. However, for parameter estimation,
control pulses with good fidelity are essential. Alternatively, one
may have to resort to more elaborate controls, nicknamed by us ``selective
dynamical recoupling'' (SDR), which we have shown to allow \emph{selective}
probing of the targeted noise source by factoring out uncontrollable
noise sources, such as an intrinsic $T_{2}$ decoherence due to a
white-noise (Markovian) process or due to pulse imperfections \cite{Smith2012,Alvarez2013}.

\section{\noindent Information gain for the real-time adaptive estimation
protocol\label{sub:Real-time-estimation-protocol}}

\noindent The best\emph{ total control and measurement time} $t'$
to be chosen for the next iteration of the real-time estimation protocol
are determined by the averaged information gain from the currently
available probability distribution of $x_{B}$, $p(x_{B}|,d,t)$,
assuming that \emph{experimental outcome} $d$ was obtained in the
iteration measured at $t$ \cite{Cory2012_Robust-online-Hamiltonian-learning}.

The expected probability to obtain the outcome\emph{ }$d'$ by measuring
the controlled spin-probe at time $t'$ from the current probability
distribution of $x_{B}$ is
\begin{equation}
p(d'|t',d,t)=\int p(d'|x_{B},t)p(x_{B}|,d,t)d\tau_{c}.
\end{equation}
The total information gain of a measurement at time $t'$ is given
by
\begin{equation}
U(t')=\sum_{d'}p(d'|t',d,t)U(d',t'),
\end{equation}
where $U(d',t')$ is the information gain if the measurement at $t'$
gives the result $d'$.

The information gain of an outcome, according to information theory,
is measured by the entropy
\begin{equation}
U(d',t')=\int p(x_{B}|d',t',d,t)log(p(x_{B}|d',t',d,t))dx_{B}.
\end{equation}
Then, the best time for the next control/measurement $t_{m}$ is defined
by the value that maximizes the expected information 
\begin{align}
U(t_{m}) & =max_{t'}\left\{ \underset{d'}{\sum}p(d'|t',d,t)\right.\label{eq:U(E_opt)-1}\\
 & \left.\int p(x_{B}|d',t',d,t)log(p(x_{B}|d',t',d,t))\, dx_{B}\right\} 
\end{align}

\begin{acknowledgments}
We thank L. Frydman and N. Shemesh for fruitful discussions. G.A.A.
acknowledges the support of the European Commission under the Marie
Curie Intra-European Fellowship for Career Development grant no. PIEF-GA-2012-328605.
G.K. acknowledges the ISF support under the Bikura (Prime) grant.
\end{acknowledgments}

\bibliographystyle{apsrev4-1}
\bibliography{bibliography}

\end{document}